\documentclass[sigchi]{acmart}
\def\BibTeX{{\rm B\kern-.05em{\sc i\kern-.025em b}\kern-.08emT\kern-.1667em\lower.7ex\hbox{E}\kern-.125emX}}
    
\copyrightyear{2019}
\acmYear{2019}
\setcopyright{acmcopyright}
\acmConference[CHI 2019]{CHI Conference on Human Factors in Computing Systems Proceedings}{May 4--9, 2019}{Glasgow, Scotland Uk}
\acmBooktitle{CHI Conference on Human Factors in Computing Systems Proceedings (CHI 2019), May 4--9, 2019, Glasgow, Scotland Uk}
\acmPrice{15.00}
\acmDOI{10.1145/3290605.3300854}
\acmISBN{978-1-4503-5970-2/19/05}

\begin{document}

%
\title[Friend, Collaborator, Student, Manager]{Friend, Collaborator, Student, Manager: 
How Design of an AI-Driven Game Level Editor Affects Creators}

%
\author{Matthew Guzdial}
\email{mguzdial3@gatech.edu}
\affiliation{%
  \institution{Georgia Institute of Technology}
}

\author{Nicholas Liao}
\email{nliao7@gatech.edu}
\affiliation{%
  \institution{Georgia Institute of Technology}
}

\author{Jonathan Chen}
\email{jonathanchen@gatech.edu}
\affiliation{%
  \institution{Georgia Institute of Technology}
}

\author{Shao-Yu Chen}
\email{shao-yu.chen@gatech.edu}
\affiliation{%
  \institution{Georgia Institute of Technology}
}

\author{Shukan Shah}
\email{shukanshah@gatech.edu}
\affiliation{%
  \institution{Georgia Institute of Technology}
}

\author{Vishwa Shah}
\email{vishwashah@gatech.edu}
\affiliation{%
  \institution{Georgia Institute of Technology}
}

\author{Joshua Reno}
\email{jreno@gatech.edu}
\affiliation{%
  \institution{Georgia Institute of Technology}
}

\author{Gillian Smith}
\email{gmsmith@wpi.edu}
\affiliation{%
  \institution{Worcester Polytechnic Institute}
}

\author{Mark O. Riedl}
\email{riedl@cc.gatech.edu}
\affiliation{%
  \institution{Georgia Institute of Technology}
}

%
\renewcommand{\shortauthors}{Guzdial et al.}

%
\begin{abstract}
Machine learning advances have afforded an increase in algorithms capable of creating art, music, stories, games, and more. 
However, it is not yet well-understood how machine learning algorithms might best collaborate with people to support creative expression.
To investigate how practicing designers perceive the role of AI in the creative process, we developed a game level design tool for \textit{Super Mario Bros.}-style games with a built-in AI level designer.
In this paper we discuss our design of the \textit{Morai Maker} intelligent tool through two mixed-methods studies with a total of over one-hundred participants. 
Our findings are as follows: (1) level designers vary in their desired interactions with, and role of, the AI, (2) the AI prompted the level designers to alter their design practices, and (3) the level designers perceived the AI as having potential value in their design practice, varying based on their desired role for the AI.
\end{abstract}

%
%


%

%
\maketitle

\section{Introduction}
Advances in Artificial Intelligence (AI) and Machine Learning (ML) systems have lead to an increasing number of people interacting with these systems on a daily basis, for example with curated social media timelines, voice user interfaces and agents, and self-driving cars. 
While the majority of these extant AI systems cover rote or repetitive human interactions, there has been increased interest in using AI to assist creative expression \cite{zhuexplainable,oh2018lead}. 
AI has been proposed in a collaborative framework in domains like storytelling, visual art, dance, and games \cite{martin2017improvisational,gatys2015neural, davis2017creative,liapis2013sentient}. 
As such approaches grow in popularity and sophistication, there is a greater need to understand how to best develop interaction paradigms  \cite{winograd2006shifting,ren2016rethinking,farooq2017human}. 
It is not yet clear how best to design interfaces and AI to support creativity. 

AI has long been utilized in video game design and development \cite{yannakakis2018artificial}.
This application of AI to game design is called procedural content generation (PCG), in which algorithms generate game content like game art, levels, and rules \cite{togelius2011search,hendrikx2013procedural}. 
In most cases, these algorithms are applied in one of two ways: (1) the algorithm is run once and a human designer then refines the output (e.g. an initial generated terrain adjusted by human designers) or (2) the algorithm is integrated into a game and produces content at runtime without designer oversight. 
Much less frequently, game design tools are developed with AI agents as an equal design partner. 
Given this, a firm understanding of the best way to structure this interaction and the consequences of employing it does not yet exist.

We designed a tool: \textit{Morai Maker}, in which a level designer can collaboratively work with an AI agent to build a \textit{Super Mario Bros.}-like platformer game level. 
The interaction occurs in a turn-based  manner, in which the human and AI designers take turns making changes to an initially blank level within the same level editor interface. 
We ran two mixed-methods studies with a total of over one-hundred participants, the first as a means of guiding the development of the tool, and the second as a means of examining the ways in which the tool impacted the behavior of practicing game designers. Based on the results of these studies we found support for the following:
\begin{itemize}
 \item The level designers varied in their desired interactions with the AI and role of the AI in those interactions. We summarize these roles as friend, collaborator, student, or manager. This followed from the designer's own artistic style, how they attempted to employ the AI, and their reactions to the AI's behavior.
 \item The level designers demonstrated a willingness to adapt their own design practice to the AI and to the level editor.
\item The level designers expressed a perception that the AI could potentially bring value to their design practice. However, the form this took depended on the individual designer's expectations for the AI.
\end{itemize}

We discuss the design implications for user interfaces and AI agents for game level editors in order to afford the most effective user-AI interactions. The main contributions of this work to the HCI community are as follows:
\begin{itemize}
 \item We discuss the design of the intelligent level design editor, both the front-end interface and back-end, collaborative deep neural network AI.
 \item Through our two mixed-methods studies we identify user behaviors and user-AI interactions in our editor.
  \item Finally, we discuss the implications for similar interfaces in which users and AI collaborate on a single creative object.
\end{itemize}

\section{Related Work}

We review works on (1) general human-AI cooperative creation or co-creation on creative tasks and (2) relevant prior work in the field of computer games.

\subsection{AI Co-Creation}

Co-creation or mixed initiative practices are those practices in which two agents work together in some creative task~\cite{horvitz1999principles}. Most typically these agents are humans, for example children working together to write a story \cite{rubegni2018design} or coworkers working on a photo collage \cite{lu2016let}. In this paper we focus on co-creation with an AI partner in the context of game level design.

The majority of recent AI advances have arisen due to deep learning \cite{lecun2015deep}, a particular machine learning approach that allows for highly complex models \cite{silver2016mastering}. There has been a significant amount of work in applying deep learning to a variety of tasks, including writing \cite{bowman2015generating,sutskever2011generating}, music composition \cite{choi2016text,huang2016deep}, and visual art \cite{champandard2016semantic}. However these prior instances almost all involve a use case in which a human user prompts a trained AI agent for output. In other words, there is only the most minimal collaboration, the AI is employed as a tool. In those cases in which the AI is not a tool, it is still not frequently given a partner role, for example being tasked with question-answering in a visual chatbot \cite{das2017visual}.

Recent work has looked into building interfaces and frameworks to allow for co-creation between a user and an AI. Sketch-RNN \cite{ha2017neural} is an interface for sketching in which an AI attempts to finish a user's drawing and generate similar drawings. Drawing Apprentice \cite{davis2015drawing} is a sketching tool that focuses on adding content based on an AI's understanding of a user's sketch. This tool allows a user to give explicit feedback to the AI, while we instead opt for implicit feedback. DuetDraw \cite{oh2018lead} is an intelligent tool for creating colored pieces of digital art. It allows two different modes with unique human-AI dynamics, with human or AI taking the leading role and the other taking an assistant role. We did not explicitly structure such roles into our models, but still found that users naturally projected different roles onto the same AI and took corresponding roles. Viewpoints AI \cite{jacob2018creative} is an intelligent dance partner installation in which an intelligent dance partner attempts to adapt its performance to its human partner.

\subsection{Games}

The concept of co-creation via procedural content generation via machine learning (PCGML) has been previously discussed in the literature \cite{summerville2017procedural,zhuexplainable}, but no prior approaches or systems exist. Comparatively there exist many prior approaches to co-creative or mixed-initiative level design agents without machine learning \cite{yannakakis2014mixed,deterding2017mixed}. These approaches instead rely upon search or grammar-based approaches \cite{liapis2013sentient,shaker2013ropossum,baldwin2017mixed,alvarez2018fostering}. This means that there is significant developer effort required to adapt each approach to a novel game.

Procedural content generation via Machine Learning \cite{summerville2017procedural} is a relatively new field, focused on generating content through machine learning methods. The majority of PCGML approaches represent black box methods, without any prior approach focused on explainability or co-creativity. Our AI could be classified as a PCGML approach, focused on co-creativity.

\textit{Super Mario Bros.} (SMB), which we employ as our particular domain of study, represents a common area of research into PCGML \cite{dahlskog2012patterns,summerville2016super,jain2016autoencoders,snodgrass2017learning,volz2018evolving}. 
Beyond working with a human user, our approach differs from prior SMB PCGML approaches in terms of representation quality and the size of generated content. 
We focus on the generation of individual level sections instead of entire levels in order to better afford collaborative level building \cite{smith2011tanagara}. 
We make use of a rich representation of all possible level components and an ordering that allows our approach to handle decorative elements.

\begin{figure}[tb]
\centering
	\includegraphics[width=3in]{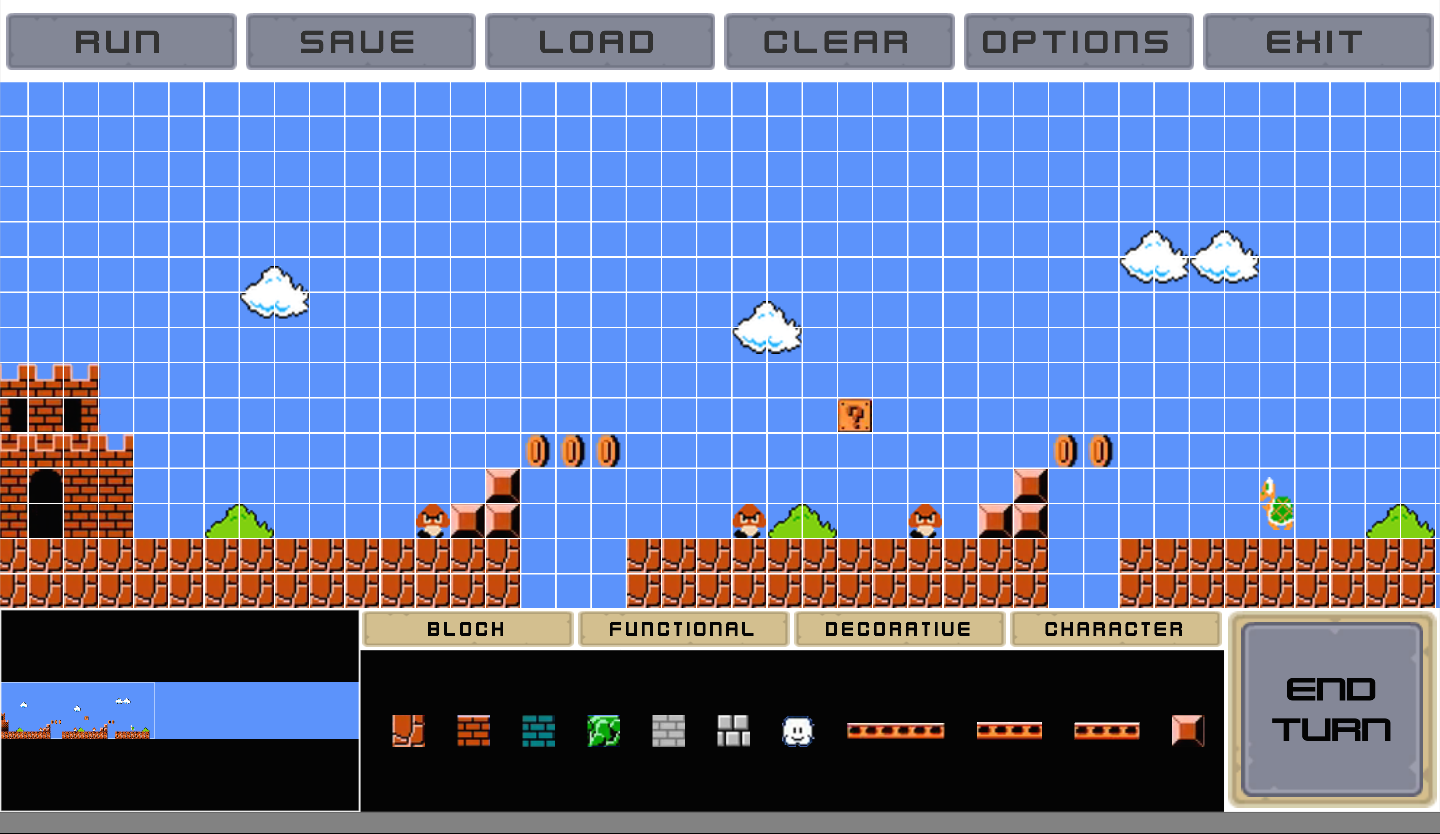}
	\caption{Screenshot of the final level editor.}
	\label{fig:screenshot}
\end{figure}

\section{Initial \textit{Morai Maker} Development}

\textit{Morai Maker} went through several iterations prior to any participant study~\cite{guzdial2016game}.
We named the initial prototype \textit{Morai Maker} after the popular level design game in the Mario franchise \textit{Mario Maker}, since it was intended as a similar application but with ``more AI".
We chose \textit{Super Mario Bros.} because there has been significant PCGML work in this area \cite{summerville2017procedural}.
Further the original game is well-recognized, which made it more likely that future study participants would be familiar with it.

We built the final version of our prototype interface in Unity3D \cite{creighton2010unity}, a game development engine.
The final version of the UI is shown in Figure \ref{fig:screenshot}. The major parts of the interface are the current level in the center of the interface, a minimap on the bottom left of the interface, a palette of sprites in the middle of the bottom row, and an ``End Turn'' button on the bottom right. By pressing this End Turn button the current AI level design partner is queried for an addition. A pop-up appears while the partner processes, and then its additions are added component-by-component to the main screen. The camera scrolls to follow each addition, so that the user is aware of any changes to the level. The user then regains control and level building continues in this turn-wise fashion. At any time users can hit the top left ``Run'' button to play through the current version of the level. A back-end logging system tracks all events, including additions and deletions and which partner (human or AI) was responsible for them. The current version of this interface Unity3D project is available on GitHub.\footnote{https://github.com/mguzdial3/Morai-Maker-Engine}

\section{Study 1: Prototype Investigation}

We ran an initial study to derive design lessons to apply to future iterations of our \textit{Morai Maker} tool, both in terms of the interface and back-end AI system. We anticipated that the choice of the AI algorithm might impact human experience. However, it is unclear how and to what degree different algorithms would impact user experience due to a lack of prior work on this particular problem. Therefore, we drew upon three previously published AI approaches to \textit{Super Mario Bros.} level generation, which we summarize in the next section. Further, we focused on a primarily quantitative user study as we were interested in deriving broad, summarizing results.  

\subsection{AI Level Design Partners}

For the initial study we created three AI agents to serve as level design partners. Each is based on a previously published PCGML approach, adapted to work in an iterative manner to fit the requirements of our editor and turn-based interaction framework. We chose to create three potential AI back-end partners because we wished to investigate what impacts each of these systems might have on the user experience. We lack the space to fully describe each system, but describe the approaches at a high level to give a sense of the means in which their behavior varied.

\begin{itemize}
\item \textbf{Markov Chain: } This approach is a markov chain from Snodgrass and Ontan{\'o}n \shortcite{snodgrass2014experiments}, based on Java code supplied by the authors. 
\item \textbf{Bayes Net: } This approach is a probabilistic graphical model, also known as a hierarchical Bayesian network from Guzdial and Riedl \shortcite{guzdial2016game}. 
\item \textbf{LSTM: } This approach is a Long Short Term Memory Recurrent Neural Network (LSTM) from Summerville and Mateas \shortcite{summerville2016super}, recreated in Tensorflow from the paper.
\end{itemize}

We chose these three approaches because they represent the most successful prior PCGML approaches in terms of breadth and depth of evaluations. 
All three systems differed in terms of the amount of existing level structure surveyed to determine what next level components to add. 
The Markov Chain looked only at a 2x2 grid of level content, making hyper-local decisions, the Bayes Net looked at a chunk of level with a width of 16 grid points, and the LSTM considered almost the entire level.
Further, the Bayes Net was the only prior approach capable of generating decorative elements and the only approach that represented each type of level component as an individual type, whereas the other two approaches grouped similar level components (e.g. all solid components represented as equivalent). We chose to only allow the agents to make additions to the level. We made this choice as the systems were designed to autonomously generate levels and so were not designed to handle deletions and to minimize the potential for the agent to undo the human's intended design of a level.

\subsection{Study Method}

Each study participant went through the same process within the same lab environment. First, they were given a short tutorial on the level editor and its function. They then interacted with two distinct AI partners back-to-back. The partners were assigned at random from the three possible options. During each of the two level design sessions, the user was assigned one of two possible tasks, either to create an above ground or below ground level. We gave the option for each participant to look at two examples of the two level types taken from the original \textit{Super Mario Bros.}.
This leads to a total of twelve possible conditions in terms of pair of partners, order of the pair, and order of the level design assignments.

Participants were given a maximum of fifteen minutes for each level design task, though most participants finished well before then. Participants were asked to press the ``End Turn'' button in order to interact with their AI partner at least once. 

After both rounds of interaction participants took a brief survey in which they were asked to rank the two AI partners they interacted with in terms of the following experiential measures:

\begin{enumerate}
    \item Which of the two agents was the most fun?
    \item Which of the two agents was the most frustrating?
    \item Which of the two agents was the most challenging?
    \item Which of the two agents most aided your design?
    \item Which of the two agents lead to the most surprising and valuable ideas?
    \item Which of the two agents would you most want to use again?
\end{enumerate}

We chose these questions to focus on particular experiential features measured in prior games experience research \cite{pedersen2009modeling,drachen2010correlation,guzdial2016game}. We employed rankings over ratings because we were particularly interested in the comparative impact on user experience and due to previously noted benefits of rankings \cite{yannakakis2015ratings}. After this ranking section, participants could chose to leave a comment reflecting on each agent. The survey ended by collecting demographic data including experience with level design, \textit{Super Mario Bros.}, the participant's gender (participants input their gender through a text box), and age (selected within the ranges of 18-22, 23-33, 34-54, and 55+).  

\begin{figure}[tb]
\centering
	\includegraphics[width=3.2in]{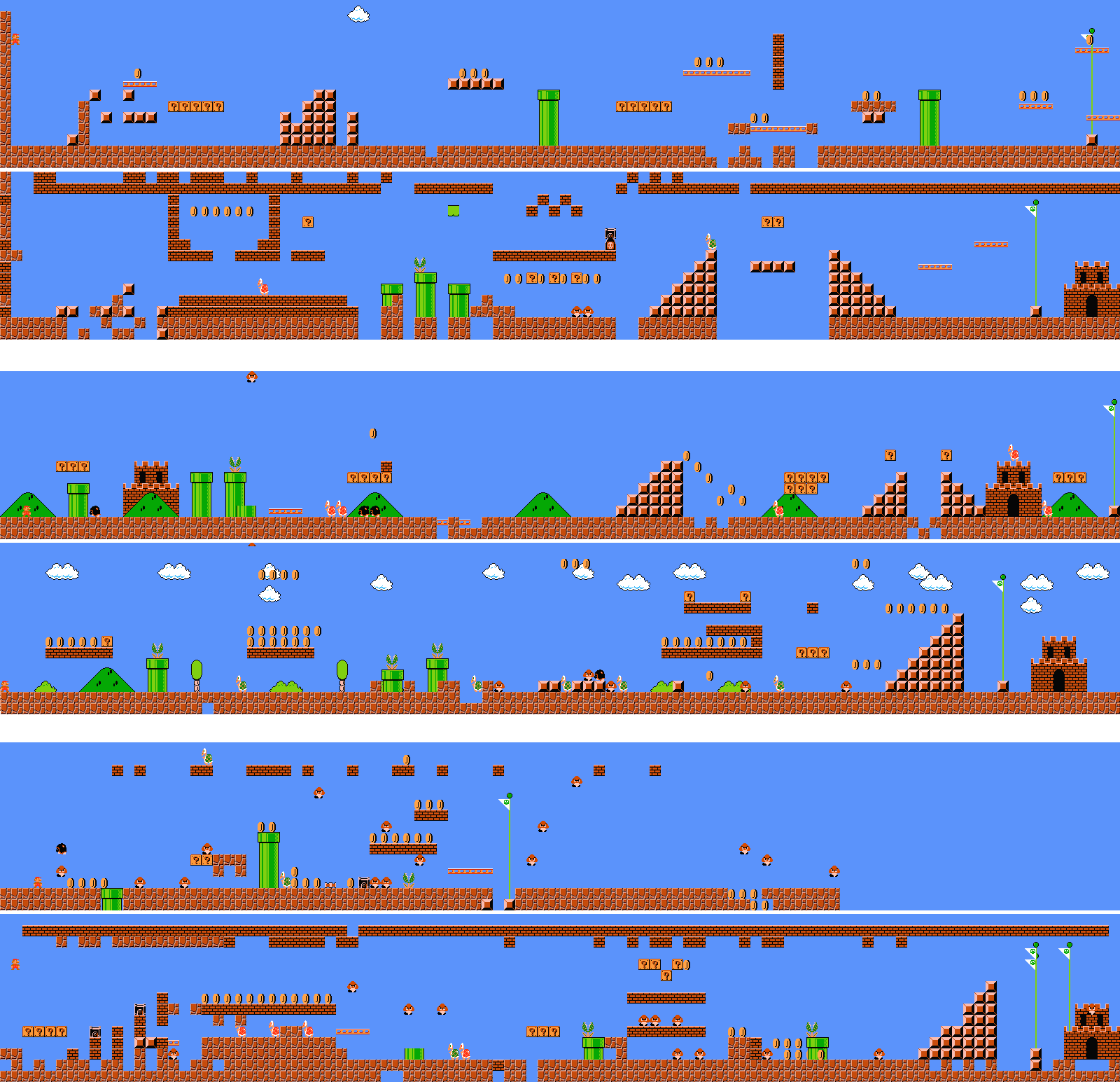}
	\caption{Examples of six final levels from our study, each pair of levels from a specific co-creative agent: Markov Chain (top), Bayes Net (middle), and LSTM (bottom).}
	\label{fig:exampleLevels}
\end{figure}

\subsection{Results}

\begin{table*}[tb]
\begin{center}
\caption{A table comparing the ratio of first rankings for the three comparisons and the $p$-value of the Wilcoxon rank-sum test, testing if the two ranking distributions differed significantly.}
\begin{tabular}{|l|c|c|c|c|c|c|c|c|c|c|c|c|c|c|} 
 \hline
 & \multicolumn{2}{|c|}{Most Fun} & \multicolumn{2}{|c|}{Most Frustrating}& \multicolumn{2}{|c|}{Most Challenging} & \multicolumn{2}{|c|}{Most Aided} & \multicolumn{2}{|c|}{Most Creative} & \multicolumn{2}{|c|}{Reuse} \\ 
 \hline
 Pair of AI & ratio & $p$ & ratio & $p$ & ratio & $p$ & ratio & $p$ & ratio & $p$ & ratio & $p$\\
 \hline
  Bayes-LSTM & 15:13 & $0.6029$  & 11:17 & $0.1142$ & 9:19 & \textbf{0.0083} & 17:11 & $0.1142$ & 19:9 & \textbf{0.0083} & 17:11 & $0.1142$\\
  \hline
  Bayes-Markov & 12:16 & $0.1469$  & 15:13 & $0.6029$ & 11:17 & $0.1142$ & 14:14 & $1$ & 10:18 & \textbf{0.0349} & 11:17 & $0.1142$\\
  \hline
  LSTM-Markov & 11:17 & $0.1142$  & 15:13 & $0.6029$ & 16:12 & $0.2937$ & 12:16 & $0.2937$ & 13:15 & $0.6029$ & 13:15 & $0.6029$\\
  \hline
\end{tabular}
\end{center}
\label{tab:subjectResults}
\end{table*}

\begin{table*}[tb]
\begin{center}
\caption{A table comparing the Spearman's correlation between the different agent ranking questions. Each cell contains the Spearman's rho and $p$ values for the ranking across the 84 participants.}
\begin{tabular}{|l|c|c|c|c|c|c|c|c|c|c|c|c|c|c|} 
 \hline
 & \multicolumn{2}{|c|}{Fun} & \multicolumn{2}{|c|}{Frustrating}& \multicolumn{2}{|c|}{Challenging} & \multicolumn{2}{|c|}{Aided} & \multicolumn{2}{|c|}{Creative} & \multicolumn{2}{|c|}{Reuse} \\ 
 \hline
   & rho & $p$  & rho & $p$ & rho & $p$ & rho & $p$ & rho & $p$ & rho & $p$\\
   \hline
 Fun & \multicolumn{2}{|c|}{-}  & -0.74 & \textbf{<2.2e-16} & -0.10 & 0.2194 & 0.79 & \textbf{<2.2e-16} & 0.76 & \textbf{<2.2e-16} & 0.88 & \textbf{<2.2e-16}\\
 \hline
 Frustrating & -0.74 & \textbf{<2.2e-16} & \multicolumn{2}{|c|}{-} & 0.21 & \textbf{0.0053} & -0.81 & \textbf{<2.2e-16} & -0.64 & \textbf{<2.2e-16} & -0.71 & \textbf{<2.2e-16}\\
  \hline
 Challenging & -0.10 & 0.2194 & 0.21 & \textbf{0.0053} & \multicolumn{2}{|c|}{-} & -0.21 & \textbf{0.0053} & -0.10 & 0.2194 & -0.07 & 0.3572 \\
   \hline
 Aided & 0.79 & \textbf{<2.2e-16} &  -0.81 & \textbf{<2.2e-16} & -0.21 & \textbf{0.0053} & \multicolumn{2}{|c|}{-} & 0.74 & \textbf{<2.2e-16} & 0.81 & \textbf{<2.2e-16} \\
   \hline
 Creative & 0.76 & \textbf{<2.2e-16} &  -0.64 & \textbf{<2.2e-16} & -0.10 & 0.2194 & 0.74 & \textbf{<2.2e-16} & \multicolumn{2}{|c|}{-} & 0.83 & \textbf{<2.2e-16}\\
  \hline
\end{tabular}
\end{center}
\label{tab:corrResults}
\end{table*}

In this section we discuss a data analysis of the results of our user study. Overall 91 participants took part in this study. However, seven participants did not interact with one or both of their partners. Thus we do not include these results in our analysis. The remaining 84 participants were split evenly between the twelve possible conditions, meaning a total of seven human participants for each condition. Of these, 67 respondents identified as male, 16 identified as female, and 1 identified as nonbinary. 64 participants placed themselves in the 18-22 age range, which makes sense given we were largely drawing from a student population, with 19 in the 23-33 age range, and 1 participant in the 34-55 age range. This population is not sufficiently diverse to draw broad, general lessons. While this fit our needs for an initial investigation, we broaden participant diversity in the second study.

62\% of our respondents had previously designed Mario levels before. This is likely due to prior experience playing \textit{Mario Maker}, a level design game/tool released by Nintendo. Our participants were nearly evenly split between those who had never designed a level before 26\%, designed a level once before 36\%, or had designed multiple levels in the past 38\%. All but 7 of the participants had previously played \textit{Super Mario Bros.}, and all the participants played games regularly. 

Our first goal in analyzing our results was to determine if the level design task (above or underground) mattered and if the ordering of the pair of partners mattered. We ran a one-way repeated measures ANOVA between ordering of agents, the pair of agents, and the design task and the experiential feature rankings. We found that no variable lead to any significance. 
Thus, we can safely treat our data as having only three conditions, dependent on the pair of partners each participant interacted with.
However, this also indicates that none of the agents were a significant determiner of the final rankings.

We give the overall ranking results in Table 1. 
We split the results into three conditions, based on the pair of agents those participants interacted with. 
This is necessary given the pairwise ranking information. 
For each pair of agents, and for each experiential feature (e.g. ``Most Fun"), we give the ratio by which each agent was ranked first for that feature and the $p$-value from running the Wilcoxon rank-sum test. 
This $p$-value is bolded in the case in which we can reject that these two sets of rankings arose from the same distribution.
To run the Wilcoxon rank-sum we represented all first place rankings as 1.0 and all second place rankings as -1.0, but we could have chosen any two distinct values.
We applied the Wilcoxon rank-sum test as our data did not fit a normal distribution.
Thus, for example for the Bayes-LSTM condition the Bayes agent was ranked as the most fun by 15 of the 28 participants while the remaining 13 ranked the LSTM as the most fun.
Note that we simplify the fifth ranking question in our survey (surprising and valuable) to most creative; surprise and value have been identified as two of the primary requirements for something to be labelled creative \cite{boden2004creative}.

The results in Table 1 indicate that we were unable to find any significant difference for all but three of the comparative rankings. 
One might suggest that we needed to collect more training data. 
However, after 72 participants we found nearly these exact same proportionate ratios. 
Instead, it may be that the participants of this study were choosing their rankings at random. 
To determine this we ran a second set of evaluations testing the degree to which rankings of one experiential feature correlated with the other experiential features. 
Table 2 summarizes the results from comparing the correlations of the experiential feature rankings. 
We applied Spearman's rho as our correlation test, due to the fact that our ranking data does not follow a normal distribution.
The first value in each cell is the value of Spearman's rho, while the second is the $p$-value of significance, with significant values in bold. 
The amount and strength of the correlations may seem surprising, but we note that there were only two possible values (first rank or second rank, associated with 1.0 or -1.0). 

These results lend support to the notion that our participants were not ranking according to random guesses, as we would not expect to see any correlation.
Instead, we can break our results into groups in terms of experiential rankings that implied a positive experience (fun, aided, creative, and reuse) and those that did not (frustrating and challenging). 
Notably, ranking the agents in terms of which was the most challenging appears to have been the least consistent across participants. 
There is a weak positive correlation with the ranking of most frustrating and a weak negative correlation with the ranking of most aided.
We anticipate this was due to a lack of clarity in how we phrased the question, but also note that some participants may have used a challenging ranking to denote a lack of understanding on their part for a potentially useful agent. 

\subsection{Study 1 Output Levels}

We give examples of two randomly selected levels co-designed by each of the three agents in Figure \ref{fig:exampleLevels}. While we chose these levels at random, they demonstrate some consistent features we found across all levels co-designed by these agents. For example, the Markov Chain agent co-designed levels (top of Figure \ref{fig:exampleLevels} have a variety of unique patterns of blocks, which do not appear in Mario or in the other agent's co-designed levels. The Bayes Net co-designed levels were more likely to contain decoration, but otherwise appear similar to typical \textit{Super Mario Bros.} levels. Finally, the LSTM agent's co-designed levels varied widely. In the cases where participants followed typical \textit{Super Mario Bros.} level conventions, the LSTM agent tended to co-design levels that strongly resembled original \textit{Super Mario Bros.} levels. However, when participants did not follow these conventions (as in the two images we randomly selected) the output tended to appear noisy or random. Despite referring to some of these levels as more or less \textit{Super Mario Bros.}-like we note that in all of these example images, and in fact in all of the final levels from this study, there existed structures that did not exist in the original \textit{Super Mario Bros.}-levels. For example in the very top level image, the extra-long orange bars and single floating cloud the use of an ``M" shape of floating blocks in the second image.

\subsection{Study 1 Results Discussion}

The lack of a consistent ranking between agents according to the experiential feature rankings suggests that no one agent stands out as significantly superior as a creative partner. Instead, they suggest that individual participants varied in terms of their preferences. This matches our own experience with the agents. When attempting to build a very standard \textit{Super Mario Bros.} level, the LSTM agent performed well. However, as is common with deep learning methods it was brittle, defaulting to the most common behavior (e.g. adding ground or block components) when confronted with unfamiliar input. In comparison the Bayes Net agent was more flexible, and the Markov Chain agent more flexible still, given its hyper-local reasoning. This can be most clearly seen in the ``Most Creative" results in Table 1, in which the Bayes Net was ranked as significantly more surprising and valuable than the LSTM, and the Markov Chain agent ranked as significantly more surprising and valuable than the Bayes Net. Despite this, in comparing between the LSTM and Markov Chain, no such significant relationship was found. 

The lack of a single superior agent is further supported by the optional comments left by some participants. For example, despite receiving the fewest number of first place rankings, some participants referred to the LSTM as ``Pretty smart overall... [collaborating] well with my ideas'', ``seemed more creative. I did actually use some of its ideas'' or felt that the agent ``seemed to build toward an idea''. 
Comparatively the participant who gave that latter quote felt the Markov agent ``didn't really seem to offer any `new' ideas'', despite the Markov agent consistently ranking higher on the question about surprising/valuable ideas. 
Notably, despite the requirement of ranking, our results indicate that many participants remained unsatisfied by both agents. One piece of evidence for this is that on average participants deleted 60\% of all the agent's additions. 
This is further made clear through comments in which participants expressed displeasure with both agents they interacted with, such as ``They both included stuff... which seemed pretty unhelpful".
One participant who interacted first with the Bayes Net agent stated that the agent ``did not pick up on my style of design" and of the second agent they interacted with, the LSTM agent, they stated they were ``Disappointed with the lack of creativity of level generation idea(s)". 
This quote appears to state the opposite view of the participants above who specifically mention the creativity of the LSTM agent's ideas.
The LSTM agent did not change between participants, thus it appears that this friction arose due to differences in the participants as level designers.
This type of friction appeared throughout the optional quotes, indicating that different participants had vastly different design values, with no one agent addressing them all.

Our primary takeaways from this first study were as follows:
\begin{itemize}
    \item Users of this prototype tool varied by such a degree that no one static agent could meet all of their expectations for an AI partner. However, users of the tool overall expressed a clear sense of what they found valuable.
    \item Participants demonstrated a consistent departure from typical \textit{Super Mario Bros.} structure. 
    \item Participants demonstrated a lack of clear understanding of their AI partners, but a willingness to invent an explanation for how these partners behaved. 
\end{itemize}

\section{Study 2: Thinkaloud}

The purpose of our first study was to derive design insights we might use to develop our initial prototype into a more fully realized tool.
Thus, we made changes based on the results of this first study, both to the interface and AI agent. 
We discuss these changes below.
After these changes we felt we had arrived at something like an ``alpha" build, to borrow software development parlance. 
We ran a second study in order to evaluate the impacts that this version of \textit{Morai Maker} had on designer's experiences and behaviors. 
This study was not meant as an evaluation of this tool, but an investigation of the effect that the tool had on users towards its continuing, future development.
We focused this study on practicing, published game designers with a more qualitative methodology for this purpose.
Before the study we identified three major research questions, which directed the design of the study and our analysis of the results. 
They are as follows:

\begin{itemize}
    \item \textbf{RQ1: }By leveraging active learning to adapt the AI partner to a user, can our tool better serve the needs of level designers?
    \item \textbf{RQ2: }Can Explainable AI allow users to better understand the AI, and therefore to better utilize the tool?
    \item \textbf{RQ3: }Will our overall changes to the tool lead to beneficial experiences for the designers?
\end{itemize}

RQ1 arose from the first and second takeaways listed in the Study 1 Results Discussion section, RQ2 arose from the third takeaway, and RQ3 arose as a general response to our changes.

\subsection{Changes to \textit{Morai Maker}}

The results of the first study indicate a need for an approach designed for co-creative PCGML instead of adapted from autonomous PCGML. In particular, given that none of our existing agents were able to sufficiently handle the variety of human participants, we expect instead a need for an ideal partner to more effectively generalize across all potential human designers and to adapt to a human designer actively during the design task. We modeled the interaction as a semi-Markov Decision Process (SMDP) with concurrent actions (the different possible additions) \cite{rohanimanesh2003learning}. Our final agent trained on the interactions with the 91 participants, using the ``Reuse" ranking (1 or -1) as the final reward, given that we felt this was the most important experiential feature and correlated strongly with the other positive features. In addition, we include a small negative reward (-0.1) if the human deletes an addition made by the AI partner.

From our human participant study we found that local coherency (Markov Chain) tended to outperform global coherency (LSTM). Thus for a proposed co-creative architecture we chose to make use of a Convolutional Neural Network (CNN) as the agent in our SMDP. We made this choice as CNNs focus learn small, local features that are helpful in making more global decisions. We made use of a three layer CNN, with the first layer having 8 4x4 filters, the second layer having 16 3x3 filters, and the final layer having 32 3x3 filters. The final layer is a fully connected layer followed by a reshape to place the output in the form of the action matrix (40x15x32). Each layer made use of leaky relu activation. We made use of a mean square loss and adam as our optimizer, with the network built in Tensorflow \cite{abadi2016tensorflow}. 

We updated our agent to use an active learning framework \cite{settles2012active}. 
During use of the tool the agent trains on implicit feedback from the user. If a user keeps the AI's additions the agent receives a reward of +0.1, and if the user removes them -0.1. 
Notably these rewards are local to the addition's placement, meaning that a user deleting a ground tile in a particular location will not impact the likelihood of seeing other ground tiles elsewhere.
We also kept track of all deletions from the user of the AI's additions in a particular session and prohibit the AI partner from making those same additions again.

Overall this had the effect of an agent that, in informal tests, was capable of picking up on user's preferences for local level structures. We predicted that this would allow us to address the first two takeaways from the results of the first study, given the ability to adapt to an individual user and training on more levels than existed in the original \textit{Super Mario Bros.}. For further technical details please see \cite{guzdial2018co}.

The size and processing requirements of this new AI model meant we could no longer run the model on the same device as the front-end editor. Instead we made use of a client-server framework, in which the AI agent ran on a server during a single participant's study, and the ``End Turn'' button prompted the server for additions. Study personnel could monitor the output of the server during the study.

Results in the first study indicated the participants did not fully understand their AI partners, to address this we included an explanation generation system. 
\textit{Explainable AI} represents an emerging field of research \cite{biran2017explain}, focused on translating or rationalizing the behavior of black box models. To the best of our knowledge, this has not been previously applied to PCGML. 
Codella et al. \shortcite{codella2018teaching} demonstrated how explanations could improve model accuracy on three tasks, but required that every sample be hand-labeled with an explanation and treated explanations from different authors as equivalent.
Ehsan et al. \shortcite{ehsan2017rationalization} made use of explainable AI for automated game playing. Their approach relies on a secondary machine learning interpretation of the original behavior, rather than visualizing or explaining the original model as our approach does. 

In our explanation generation system we identify the most decisive 4x4 slice of the level for a particular addition as a printout from the server, we also included the AI agent's confidence in the addition (based on the activation) and the filter of the first layer of the CNN that was maximally activated as described in \cite{olah2018building}. Essentially this meant that for every addition an AI expert could translate why the AI made that addition. We did not incorporate explanations into the editor at this time, given a lack of clarity on how best to integrate this feature. 
Instead we treated this as a semi-Chinese room \cite{searle1999chinese} or semi-Wizard of Oz \cite{riek2012wizard} design, with study personnel acting to translate the output explanation into user-understandable language.
For consistency, we used a single member of the study personnel to translate these explanations. 
No hypothesizing was included in these explanations, only relating the output in terms of the generated explanation and prior experience. 
Our hope was that we might use the ways in which this individual translated the explanations as insight into an interface-integrated explanation system.

There were only two changes made to the front-end level editor. First, we replaced the ``Options'' button with a ``Remove" button that would remove all the AI's additions from the last turn. We made this choice because we anticipated the potential for brittleness and incoherent additions prior to the agent successfully adapting to the user. The second change was to remove the functionality of the ``Run" button, since in the case of the 7 participants who did not make use of the AI in the first study, they spent their full time playing their own level.

\subsection{Method}

The overall method of this second study remained similar to the first.
The study differed in terms of the location.
Our goal for this study was to determine the impact of this tool on user's design practice. 
For this reason, we gave the option for study participants to take part in the study wherever they felt the most comfortable to engage in level design. 
In the cases where users took part in the study remotely, study personnel used conference software to watch their screen and record audio.

At the start of the study, participants were given a full run-down of the tool, this time including informing the users they could ask questions to the study personnel, including explicitly ``explanations about the AI". 
Participants then took part in designing two levels. 
For each level they now interacted with the same agent, which was not reset in between levels.
Thus the agent would ideally continue to adapt to the designer during the course of the study session, and we would expect to see better performance in the second level design session.
The participants were given a random level design task for each level, to create an above ground or under ground level, while the AI was not altered in either case.
The major difference in the level design portion of the study was the use of a think-aloud protocol.
At the beginning of the study session participants were encouraged to voice their ``thoughts, reactions, and intentions" aloud.
In the case where the participant remained quiet, study personnel asked ``why'' questions, referring back to the previous actions of the participant. 
The exact choice of when to make use of these questions was left up to individual study personnel. 
A single study personnel ran each of these sessions.
We made this choice for consistency between when the ``why" questions were asked and the generated explanations.

Given that the original survey was designed to compare between interactions with two different agents, we reworked it for interacting with the same agent twice. In particular we reworded four of the original questions to ask participants to rank between the two experiences, in regards to the agent. We did not include the question on reuse, as it was not relevant, and we did not include the question on which agent was most challenging to use given the apparent issues with that question.

We added a second set of questions after the initial experiential feature rankings meant to specifically address our research questions. They were as follows:

\begin{enumerate}
    \item Did you prefer the agent's behavior in the first or second session? (a) First (b) Second.
    \item Would you prefer to use this tool with or without the AI partner? (a) With (b) Without (c) No preference.
    \item Did you feel that the agent was collaborating with you? (a) Yes (b) No
    \item Did you feel that the agent was adapting to you?  (a) Yes (b) No
    \item If you asked for explanations, did you find that they improved your experience?  (a) Yes (b) No
\end{enumerate}

For each of these questions participants were asked to explain their answers in text, except for the last one, which asked for an explanation in the case of a positive response. 
These questions were intended to help us pick apart elements of user experience related to our two research questions, RQ1~(1, 4), RQ2 (5), and RQ3 (2, 3).
We added questions to the demographic section to address the particular goals for this study. We asked the respondents to describe their game design experience (offering hobbyist, industry, indie, academic, etc. as suggested answers), and asked them to rate their experience with with AI/ML on a Likert scale (No experience, I have studied or used AI/ML in the past, I regularly use AI/ML, I am an AI/ML expert). We asked this question due to RQ3, and our concern that experience with AI may impact how participants asked for explanations.

\begin{figure}[tb]
\centering
	\includegraphics[width=3.2in]{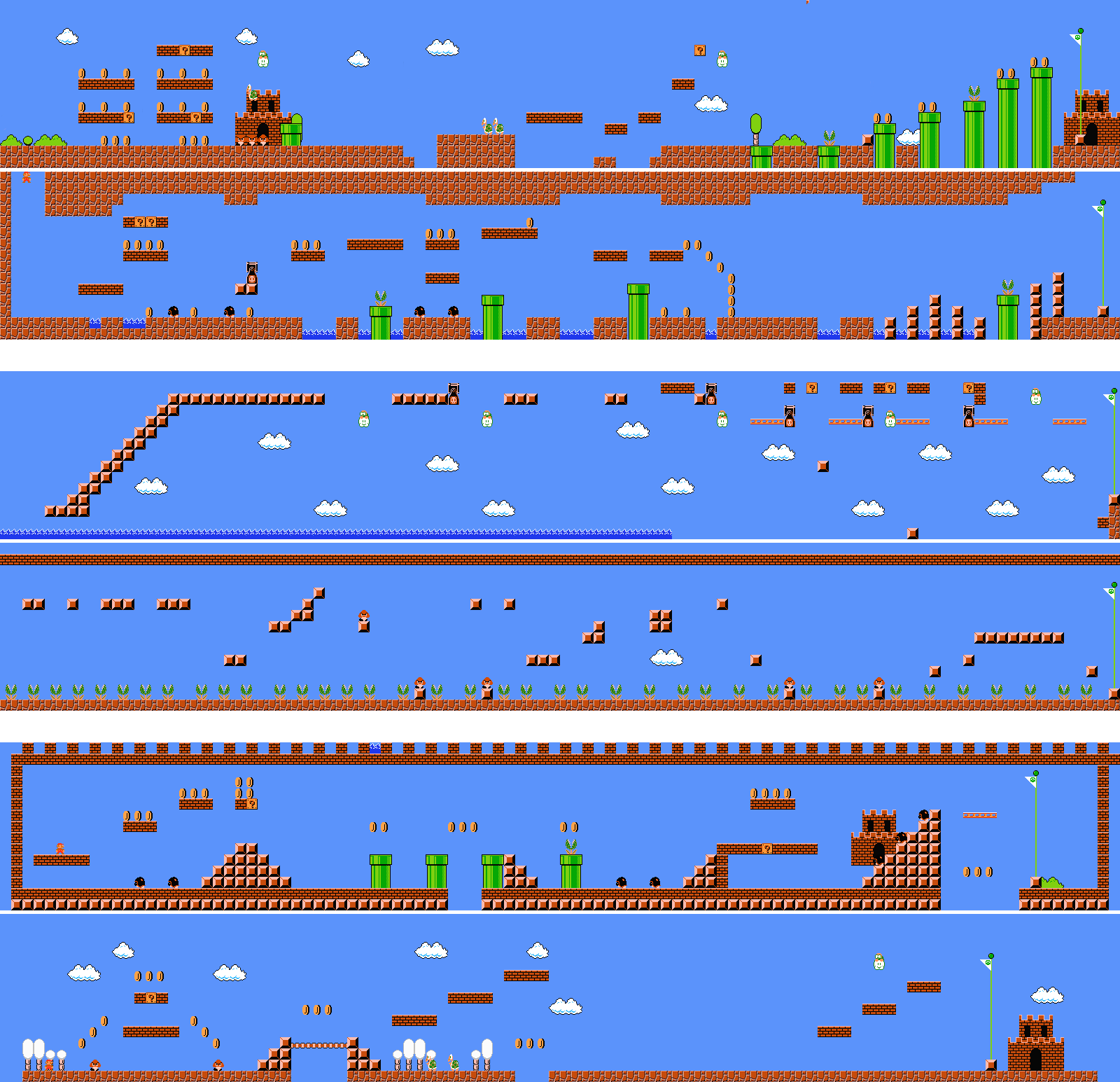}
	\caption{Examples of six final levels from our second study, each pair of levels comes from a single participant.}
	\label{fig:study2Levels}
\end{figure}

\subsection{Analysis Methods}

During the think-aloud design portions of each study personnel were instructed to note interactions or utterances they felt related to our three research questions. After all participant data had been collected we further reviewed the audio recordings for any additional interactions or utterances. We engaged in two discussion sessions between members of the team concerning this final set of notes in order to derive our final discussion points in terms of our research questions, grouping each noted utterance or interaction in terms of whether it positively or negatively related to each research question. 

\subsection{Demographic Results}

We reached out to a total of 24 game designers through various social media accounts. We sought only those with published games, either indie, hobbyist or industry. We made this choice because we were interested in designers with a consistent design practice, in order to determine how this design practice was impacted by our tool. 16 agreed to take part in this study. During the study session of two of these individuals it was discovered that networking issues (a firewall the participant had no control over and network latency) lead to the tool crashing. Thus we were left with 14 final participants.

Of the 14 participants, 8 identified as male, 4 as female, and 2 as nonbinary. While not ideal, this does outperform the games industry in terms of gender diversity \cite{igda17}. In terms of age, 11 chose 23-33, 2 chose 34-54, and 1 chose 18-22. This is notably older on average than our initial study, which follows from the difference in participant pools.

We randomly selected three of our participants and visualize their final levels in Figure \ref{fig:study2Levels}. These levels vary significantly from the types of levels seen in Figure \ref{fig:exampleLevels}, and each demonstrates unique design aesthetics. Most notably the middle designer, who described her levels as ``sadistic''. These levels offer some evidence towards the tool supporting creative expression, but does not reflect the interaction with the AI. 

\begin{table}[tb]
\begin{center}
\caption{Ratio of the answers to the survey's question in the second study.}
\begin{tabular}{|l|c|c|} 
 \hline
 & First & Second\\
 \hline
 Most Fun & 5 & 9\\
 \hline
 Most Frustrating & 8 & 6\\
 \hline
 Most Aided & 5 & 9\\
 \hline
 Most Creative & 5 & 9\\
 \hline
 Preference & 6 & 8\\
 \hline
 \hline
 & Yes & No\\
 \hline
 Collaborating & 7 & 7\\
 \hline
 Adapting & 9 & 5\\
 \hline
\end{tabular}
\end{center}
\label{tab:ratioResults}
\end{table}

\subsection{Quantitative Results}

The only quantitative results of this study are the survey responses and the logging information. We summarize the results in terms of the ratio of answers in Table 3. It does not include the answer to the question concerning whether the participant would want to use the tool with or without the AI, but we found that nine answered ``with'', two answered ``without'', and the rest ``no preference''. Since the AI partner is the main feature that differentiates this tool from other level design editors, we anticipate that this result indicates some positive support for RQ3. Overall these results indicate a majority of participants had a positive interaction with the agent, which gives some support to a positive answer to RQ1.

As a summarizing statistic we include the change in average ranking of the second agent between the four experiential features shared between the two studies. This indicates that individuals in the second study were 14.3\% more likely to rank the second design experience as more fun than individuals in the first study, 15.5\% less likely to rank the second agent as more frustrating, 20.3\% more likely to think the second experience aided their design, and 9.5\% more likely to view the second experience as demonstrating more surprising and valuable ideas from the AI. This lends further support toward a positive answer to RQ1 since by the time of the second session the AI agent should have had time to adapt to the user. 
Half of the participants felt the AI was collaborating with them. This seemed to be due to differing expectations on what AI collaboration should look like across the participants, which we discuss further below. We ended up not finding the explanation question useful because only three of the fourteen respondents asked for any explanations having to do with the AI. Only 45\% of AI additions were deleted on average in this study, compared to 60\% in the first. However we caution against strong interpretations of these results.

\subsection{Qualitative Results}

In this subsection we organize the results of our analysis of the think-aloud utterances and design tool interactions in terms of our three research questions. In our analysis we noted two major themes across our findings. First, that individuals differed widely in terms of their expected roles for the AI. Second, a willingness to adapt their own behavior to the AI. We identify participants according to the order that they took part in the study.

Our first research question asked \textbf{by leveraging active learning to adapt the AI partner to a user, can our tool better serve the needs of level designers?}. Overall, we found that the AI consistently demonstrated adaptation to the participants. The study participants indicated that they noticed this, both during the think-aloud and in their comments on the final survey. Study participant 9, a male hobbyist game designer, added this after the collaboration survey question: ``The more I placed, the more the agent seemed to get a sense of what I was going for, and in the end we had a couple of decent Mario levels." Participant 1, a male indie game designer, responded to the question about round preference with: ``... After I rejected some of the off-theme decorative elements it attempted to add at the beginning, I felt like it recognized what kind of blocks I was working with in this environment." During the study, every participant had at least one interaction where they noted the AI's adaptation. Participant 5, a female academic game designer reacted to one of the AI's additions with ``This is a little better...Hey it's starting to figure me out a little bit". Participant 6, a non-binary, hobbyist game designer offered the following quotes directly after observing each round of the AI's additions ``Still don't feel quite inspired? I'll try a tighter feedback loop", ``Wait? that was nice", ``Not great, except this part is really great", and ``Honestly, good job AI?".

Each participant experienced and remarked upon at least one instance of the AI adapting to them. However, for some participants, this was the exception, with most of the behavior appearing ``random". On a comment after the first/second round behavior preference question, participant 14, a hobbyist nonbinary game designer stated ``The second agent placed objects fairly arbitrarily, in places where it didn't really affect gameplay, just looked weird". Participant 11, a male game designer, who described his game design career as ``all of the above except industry" stated ``it learned too much from the first and then attempted to apply it to the second". Notably this was one of the three participants who actually asked for explanations about the AI agent. 

Our second research question asked \textbf{can Explainable AI allow users to better understand the AI, and therefore to better utilize the tool?}. We did not find a meaningful answer to this research question. Only three participants asked for explanations, with two responding to the survey saying it was useful. Participant 7, a male hobbyist game designer added the comment ``It helped me understand what the AI was attempting to do and adapt to it for better collaboration." However, this could have been prompted by the way we phrased the survey question. Instead of asking for explanations about the AI, almost all participants instead voiced hypotheses about why the AI did what it did. In cases where these were phrased as questions (e.g. ``Will it learn?" from participant 5), our study personnel asked if this was meant as a question, to which our participants responded in the negative. This came up repeatedly, even with participants who asked for explanations. Participant 8, a male who described himself as a ``mostly" hobbyist, stated ``It is adding something to help the players", despite asking whether the system had any explicit model of the mechanics and hearing it did not. These hypothesized reasons often anthropomorphised the AI as in ``I'm happy with what the AI is thinking here" from participant 11 and ``I like where it's heads at but- I'm gonna trim some things" from participant 13, a female student/indie game designer.

Our third research question asked \textbf{will our overall changes to the tool lead to beneficial experiences for the designers?}, by which we address our belief that designers would find value in the tool. First, participants generally praised the front-end interface. Participant 10, a female ``industry, indie, academic" game designer stated ``Your level editor is amazingly functional for something written in Unity. Also, it felt pretty good to use." Across the think-aloud and survey comments we identified two major strategies to get value from the tool. As either an unintentional inspiration source or an intentional means of getting over a lack of ideas. Participant 4, a male hobbyist designer indicated that he preferred the behavior of the first experience to the second, explaining ``It added things that made the level harder, things that I would not have thought of by myself, most likely". Participant 6, after stating a preference for the tool with the AI partner stated ``I really like the tool regardless of the AI partner, but it was nice to be surprised by the AI partner! It prompted conversation/discussion in my head". Participant 14 during the think-aloud stated ``I'm running out of ideas", prompted the AI for additions, and exclaimed ``Oh yeah I forgot about these things!". Participant 5 stated she would prefer to use the tool with the AI and explained ``It was more fun than facing an empty level by myself". 

Not everyone found the tool to be consistently valuable, with the majority of complaints about the tool focusing on the AI's behavior. Participant 10 indicated she had no preference about whether to use the tool with or without the AI and stated ``I could see using this tool as a way to give myself inspiration. But if I had more specific goals in mind... I would have found it more inhibiting than useful". Seven of the participants used the term ``random" to refer to behavior they disliked from the AI, as with participant 2, a female academic game designer, who commented ``the AI choices felt a bit too random to suit my taste", after indicating she'd rather use the tool without the AI. Participants also had issues with the AI not recognizing intentional empty space. Even though the AI could not make the same additions in the same locations, it could add the same additions in other map locations. When the AI repeatedly filled Participant 11's gaps with ground blocks he exclaimed ``Quit filling in the gaps!". Some users expected the AI to only add content that a potential player could reach given the current level. Participant 9, reacting to an AI addition stated ``that's an unreachable block so I'm going to delete it", this was a common reason given for deleting the AI's additions.

\subsection{Roles and User Adapting Analysis}

Users differed widely in their expectations for their AI partner.
In our analysis we identified four major expected AI roles: friend, collaborator, student, and manager. 
These expected roles could positively or negatively impact the user experience, and fluidly changed throughout each design session.
For friend we indicate those participants who viewed interaction with the AI as primarily a fun activity, and even literally described the AI as a friend. 
Participant 13 began her second design session by clicking the end turn button and stating ``Let's see what my friend comes up with". 
Participant 7 described the experience as ``fun", but gave the answer of no preference in the question of whether he would prefer the tool with or without the AI. 

Collaborator as a role indicated those who wanted an equal design partner. 
This could be positive, for example participant 2 stated she ``found myself wanting to compliment the AI's choices more as I noticed its responsiveness during the second level build". 
It could also be negative, when the AI did not match a participant's expectations for a human collaborator. 
Participant 11 stated ``What I expect from a design partner is one contributing complete(ish) ideas rather than small edits" after answering that he did not think the AI collaborated with him. 
Participant 2 echoed these statements, but with a different sense of collaboration: ``It didn't seem to be trying to create any consistency with my initial choices as you might expect from a human collaborator".

For the student role, we indicate that the participant seemed to expect the AI to follow their specific design beliefs or instructions. Participant 8 gave ``I think that the first agent give less `illegal' suggestions" as his reasoning for preferring the first agent, suggesting a notion of right and wrong agent behavior. On a more positive end, participant 7 stated he felt the system was collaborating with him, explaining ``It tried to copy things I did and it wasn't `frustrating' to work with." For this participant, copying him was the ideal for a collaborator. Lastly, perhaps with a cynical view of a student as underling, participant 14 reacted to AI additions by exclaiming ``Yes! Do my work for me!"

By manager, we indicate that the participant seemed to view the AI as giving instructions to them or judging their design. 
Many of the cases of participants adapting their behavior to the AI occurred within this interactive framework.
Participant 4 began his session by asking if it was possible for the AI to ``evaluative" him.
Later, reacting to an AI addition he stated ``so I like what the AI- I'll follow along with what it suggested". This notion of ``following along" with the AI was common across participants who engaged with the AI in this way. As with Participant 2, who reacted to an AI addition by stating ``I'll stick to its idea".
Participant 8, reflecting on the AI's previous additions while designing stated ``I'll probably just add some blocks to satisfy its needs".

Some participants adapted their behavior to the AI primarily as a means of attempting to determine how best to interact with it. 
Participant 12, a male indie, paused and reflected on the AI's additions during one of his sessions, stating ``It does do pretty cool stuff now and then, I think I need to get more used to what its doing".
Others were more explicit in their plan to experiment, with participant 11 stating near the beginning of his first level design session ``[I'm going to] spend some time testing it". Still others adapted their behavior as an extension of inspiration from the AI, as with participant 6 reflecting with ``It does make me want to do more weird things". Finally, there were those who decided to attempt to change their own design to make even the most ``random" AI additions fit, like participant 8, who gave ``i am looking for ways that make [the AI additions] legit" as an explanation for changing his design.

\section{Discussion and Design Implications}

Overall, we found that our second study demonstrated clear answers for our research questions concerning adaptation of the agent and the value user's found in the tool. However, we found no clear answer for our research question concerning explainable AI helping users to leverage the tool. Instead, it seemed users were much more likely to come up with their own explanations for the AI's performance. We anticipate this could potentially have arisen from a lack of experience interacting with software and tools with explanations. Alternatively, given that we used a human to translate our Explainable AI system there may have been social pressure not to ask questions related to the AI.

For ourselves and for future designers of similar tools, we identified a few design implications or takeaways from both studies. First, users vary widely in terms of their expectations for the AI, in terms of the role the AI should take in collaboration and its performance in that role. One could propose a technical solution to this problem, that the AI should just adapt more successfully, but we expect more success to come from more explicitly outlining what the AI can and cannot do. Second, users expressed an interest in adapting to the AI, either because they viewed the AI as the more dominant designer or because they were motivated to find the best ways to interact with it. This is a positive sign for anyone interested in co-creation between AI and human designers, as it indicates a willingness to bend design practice to incorporate AI. Finally, we found that users overall found value in the tool as a source of inspiration, in terms of its ability to encourage users to rethink a design or to offer new ideas when a designer was unsure how to proceed. We anticipate that these interactions would transfer to other design domains.

\section{Potential for Negative Impacts}

With any machine learning and artificial intelligence application there is a need to interrogate the potential to take away the livelihood of people. We note that this tool was intended to function only as a design aide, not as a replacement for a designer. The AI systems described in this paper are insufficient to act as solitary designers, developed in favor of augmenting, not replacing, creative work.

Given that our AI approach adapts to the designs of a particular user, it could potentially reinforce design biases. For example, replicating a designer's use of an offensive or over-used design, such as yet another game level in which one must save the princess. There is no place for our system as it currently exists to challenge a designer to improve beyond inspiring the designer to employ additional level components. We anticipate that there is not a great risk for negative impacts in the domain of \textit{Super Mario Bros.}, but continue to reflect on this for future versions of the tool.

\section{Conclusions}

In this paper we discuss the development of an intelligent level editor, which includes an AI partner to collaborate with users. We ran two mixed-methods user studies, with the first focusing on developing the initial tool, and the second focusing on interrogating its affect on practicing level designers. We found that users varied widely in their expectations and expected role of the AI agent, demonstrated a willingness to adapt their behavior to the agent, and overall viewed the tool as having potential value in their design practice.

%
\begin{acks}
This material is based upon work supported by the National Science Foundation under Grant No. IIS-1525967. Any opinions, findings, and conclusions or recommendations expressed in this material are those of the author(s) and do not necessarily reflect the views of the National Science Foundation. We would also like to thank the organizers and attendees of Dagstuhl Seminar 17471 on Artificial and Computational Intelligence in Games: AI-Driven Game Design, where the discussion that lead to this research began. This work was also supported in part by a 2018 Unity Graduate Fellowship.
\end{acks}

%
\bibliographystyle{ACM-Reference-Format}
\bibliography{sample-sigchi}

%

\end{document}